\documentclass[runningheads]{llncs}
\usepackage[T1]{fontenc}
\usepackage{amsmath}
\usepackage{amssymb}
\usepackage{amsfonts}
\usepackage{url}       

\usepackage{graphicx}
\usepackage{float}

\begin{document}
\title{Synthetic EEG Generation using Diffusion Model for Motor Imagery Task}
\titlerunning{Synthetic EEG Diffusion Model}
%
\author{Henrique de Lima Alexandre\inst{1}\orcidID{0009-0002-1918-4041} \and
Clodoaldo Aparecido de Moraes Lima\inst{1}\orcidID{0000-0003-3381-5348
}}
\authorrunning{H. L. Alexandre and C. A. M. Lima}
%
\institute{Escola de Artes, Ciências e Humanidades (USP) São Paulo, SP
\email{\{henrique\_lima,c.lima\}@usp.br}}
\maketitle              
\begin{abstract}
Electroencephalography (EEG) is a widely used, non-invasive method for capturing brain activity, and is particularly relevant for applications in Brain-Computer Interfaces (BCI). However, collecting high-quality EEG data remains a major challenge due to sensor costs, acquisition time, and inter-subject variability. To address these limitations, this study proposes a methodology for generating synthetic EEG signals associated with motor imagery brain tasks using Diffusion Probabilistic Models (DDPM). The approach involves preprocessing real EEG data, training a diffusion model to reconstruct EEG channels from noise, and evaluating the quality of the generated signals through both signal-level and task-level metrics. For validation, we employed classifiers such as K-Nearest Neighbors (KNN), Convolutional Neural Networks (CNN), and U-Net to compare the performance of synthetic data against real data in classification tasks. The generated data achieved classification accuracies above 95\%, with low mean squared error and high correlation with real signals. 

Our results demonstrate that synthetic EEG signals produced by diffusion models can effectively complement datasets, improving classification performance in EEG-based BCIs and addressing data scarcity.

\end{abstract}

\section{Introduction}

Electroencephalogram (EEG)-based Brain-Computer Interfaces (BCI) have gained prominence due to their non-invasive nature, low cost, and applicability in fields such as neuroengineering, rehabilitation, and biometrics. However, the collection of high-quality EEG signals remains a challenge due to interindividual variability, susceptibility to noise, such as muscle artifacts, electrical interferences, and ocular artifacts, and the high cost of multichannel systems \cite{casson_2010} \cite{michel_2004}. These limitations directly hinder the creation of representative datasets, limiting the training of robust and generalizable machine learning models.

In this context, the artificial generation of EEG data has emerged as a promising solution. Techniques based on Generative Adversarial Networks (GANs), Variational Autoencoders (VAEs), and more recently, Denoising Diffusion Probabilistic Models (DDPMs), have been explored with the goal of expanding datasets and capturing the complexity of real signals \cite{penava_2023} \cite{ho_2020} \cite{song_2020}. Among these approaches, diffusion models stand out for their training stability and their ability to generate high-fidelity samples, overcoming limitations such as the mode collapse commonly observed in GANs \cite{dhariwal_2021} \cite{goodfellow_2014}.

Diffusion models operate by modeling the data probability distribution through a progressive noise addition process, followed by a reverse denoising process, enabling the generation of realistic samples from Gaussian noise \cite{sohl_2015}. This characteristic is particularly well-suited for EEG data, which exhibits complex temporal structures and high sensitivity to perturbations.

This work explores the application of diffusion models for the generation of synthetic EEG channels, with a focus on motor imagery tasks. The proposed approach is compared against Conditional Wasserstein GAN with Gradient Penalty (cWGAN-GP), evaluating the quality of the generated signals under multiple quantitative and qualitative metrics. As an additional contribution, the impact of using synthetic data on the classification accuracy of models trained in data-limited scenarios is also investigated.

\section{Related Work}

A review of recent literature has identified a significant growth in the use of generative models for the artificial generation of EEG signals, particularly for motor imagery tasks. Techniques such as GANs, VAEs, and more recently, Diffusion Models have been widely investigated to address data scarcity and improve the quality of synthetic signals used in BCI systems.

GAN-based models have been the most frequently employed, with applications ranging from topographic amplification to enhancement of the spatial resolution of EEG signals. Approaches such as those proposed by \cite{raoof2023}, \cite{zhao2023eeg}, and \cite{luo2020} demonstrated that conditional GANs and specialized variants can significantly improve classifier performance in motor imagery tasks. However, these approaches face challenges such as training instability and mode collapse.

As an alternative, Diffusion Models have gained prominence due to their ability to reconstruct EEG signals with high fidelity through a progressive denoising process. Studies such as \cite{torma2023} and \cite{qian2024} have shown that this approach more robustly preserves the temporal and spectral characteristics of biological signals.

Additionally, more recent works have proposed hybrid architectures, such as cWGAN-GP, DHCT-GAN \cite{cai2025} and NeuroDM, which integrate CNNs, Transformers, and attention mechanisms to further enhance the quality and generalization of the generated signals. Despite these advances, the literature still lacks studies specifically focusing on the generation of synthetic motor imagery signals with quantitative evaluation and comparisons across multiple classifiers.

This work aims to address this gap by proposing a DDPM approach for the reconstruction of EEG channels associated with motor imagery tasks, directly comparing it with cWGAN-GP methods and evaluating the generated signals in classification tasks using KNN, LR, CNN, and U-Net.

\section{Methodology}
\subsection{Dataset}
The dataset used in this study is the BCI Competition III – Dataset V, which contains EEG signals recorded from three healthy subjects during four sessions of motor imagery tasks without feedback. The signals were acquired at a sampling rate of 512 Hz using 32 channels positioned according to the international 10-20 system.

During the experiments, the subjects imagined movements of the left hand, right hand, feet, and tongue, corresponding to the four target classes of the study. Each session includes multiple repetitions of each task, totaling approximately 280 samples per class per subject. The data were provided in MATLAB format, containing the raw signals and their corresponding labels.

For the experiments in this work, data were split into 70\% for training and 30\% for testing. All samples were preprocessed as described in the next subsection.

\subsection{Data Preprocessing}

The EEG dataset used in this study was acquired at a sampling rate of 512 Hz, using 32 channels positioned according to the international 10-20 system. Data processing was performed using the MNE-Python library.

Initially, a Butterworth bandpass filter between 8 and 30 Hz was applied, covering the mu (8–13 Hz) and beta (13–30 Hz) bands, which are commonly associated with motor imagery tasks. Next, automatic annotation of muscle artifacts was performed based on high-frequency (20–140 Hz) z-score analysis, followed by visual inspection for ocular artifacts.

Artifact correction was conducted through Independent Component Analysis (ICA), with automatic exclusion of components related to eye movements, identified using the Fp1 and Fp2 channels as references. After artifact removal, the standard 10-20 montage was assigned for spatial visualization.

Each channel was then standardized using the z-score over time to reduce inter-subject variability. Epochs corresponding to motor imagery tasks were segmented and labeled into four classes. For the experiments involving generative models, the Fp1, Fp2, AF3, AF4, F7, F8, T7, and T8 channels were removed and synthetically reconstructed based on adjacent channels.

\subsection{Channel Selection for Synthetic Reconstruction}

In order to optimize the efficiency of the BCI system and reduce computational complexity, we selected EEG channels less relevant for motor imagery tasks—specifically, Fp1, Fp2, AF3, AF4, F7, F8, T7, and T8 —to be synthetically reconstructed. This approach is supported by previous studies demonstrating that the exclusion or reconstruction of redundant or less informative EEG channels can improve system performance without compromising classification accuracy \cite{tang2022channel,yu2020cross,alotaiby2015review}. 

Typically, the most informative EEG channels for motor imagery tasks are located over the sensorimotor cortical areas (e.g., C3, Cz, and C4), as established by the international 10-20 system. Conversely, frontal (Fp1, Fp2), temporal (T7, T8), and prefrontal (AF3, AF4) channels are often considered less relevant for these specific tasks, making them suitable candidates for synthetic reconstruction and further analysis in data augmentation strategies.

\begin{figure}[htbp]
    \centering
    \includegraphics[width=0.6\textwidth]{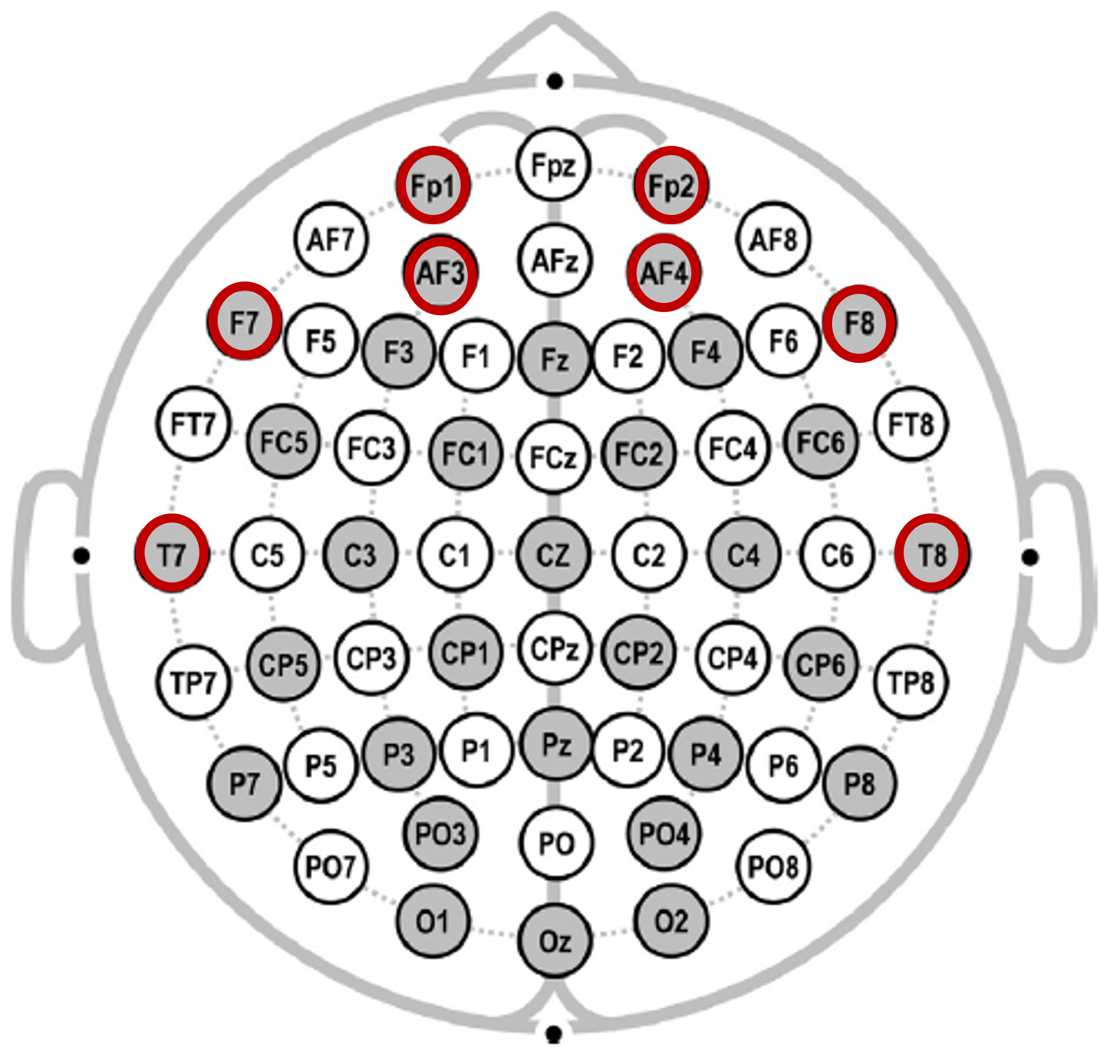}
    \caption{EEG channels selected for synthetic reconstruction (highlighted in red). These channels (Fp1, Fp2, AF3, AF4, F7, F8, T7, and T8) were chosen due to their relatively lower relevance in motor imagery tasks, based on previous literature \cite{tang2022channel,yu2020cross,alotaiby2015review}. The remaining channels represent areas typically considered highly relevant for capturing motor imagery-related neural activity.}
    \label{fig:canais_sinteticos_selecionados.png}
\end{figure}

\subsection{Diffusion Model Architecture}

In this work, we implemented a conditional diffusion model utilizing an adapted 1D U-Net architecture specifically designed for generating synthetic EEG signals. Our approach is inspired by the Denoising Diffusion Probabilistic Model (DDPM) methodology \cite{ho2020denoising}, consisting of forward and reverse diffusion processes.

The forward diffusion process involves gradually introducing Gaussian noise to the original EEG signal data $x_0$ over a defined number of timesteps $T$, specifically 1000 in this implementation. This noise schedule is defined linearly, progressing from a low variance $\beta_{\text{start}} = 0.0001$ to a higher variance $\beta_{\text{end}} = 0.02$, formally defined as:
\begin{equation}
\beta_t = \text{linspace}(\beta_{\text{start}}, \beta_{\text{end}}, T)
\end{equation}

At each timestep $t$, the noisy version of the original data $x_t$ is computed by:
\begin{equation}
x_t = \sqrt{\alpha_t} \cdot x_{t-1} + \sqrt{1 - \alpha_t} \cdot \epsilon_t, \quad \epsilon_t \sim \mathcal{N}(0, I)
\end{equation}
where $\alpha_t = 1 - \beta_t$, and the cumulative product $\bar{\alpha}_t = \prod_{i=1}^t \alpha_i$ is used during training.

\begin{figure}[htbp]
    \centering
    \includegraphics[width=0.8\textwidth]{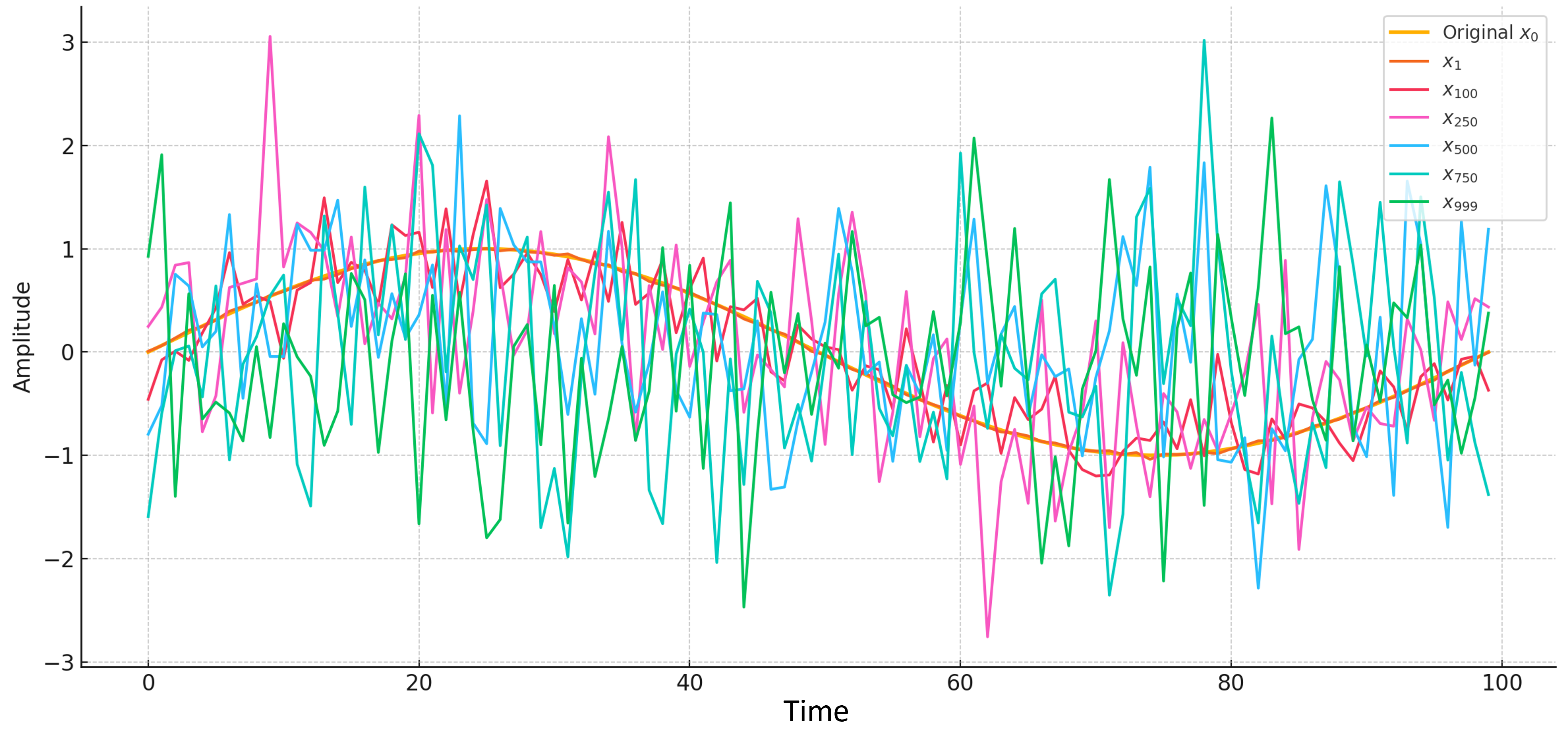}
    \caption{Forward process (Diffusion) in a Simulated EEG signal}
    \label{fig:foward_process}
\end{figure}

Our reverse diffusion (denoising) process aims to reconstruct the original signal $x_0$ from the noisy signal $x_T$ using a custom 1D U-Net architecture integrated with sinusoidal positional embeddings to effectively encode temporal information at each timestep. The reverse diffusion step can be formulated as:
\begin{equation}
x_{t-1} = \frac{1}{\sqrt{\alpha_t}} \left( x_t - \frac{\beta_t}{\sqrt{1 - \bar{\alpha}_t}} \cdot \epsilon_\theta(x_t, t) \right)
\end{equation}
where $\epsilon_\theta(x_t, t)$ is the noise predicted by our neural network at timestep $t$.

\begin{figure}[htbp]
    \centering
    \includegraphics[width=0.8\textwidth]{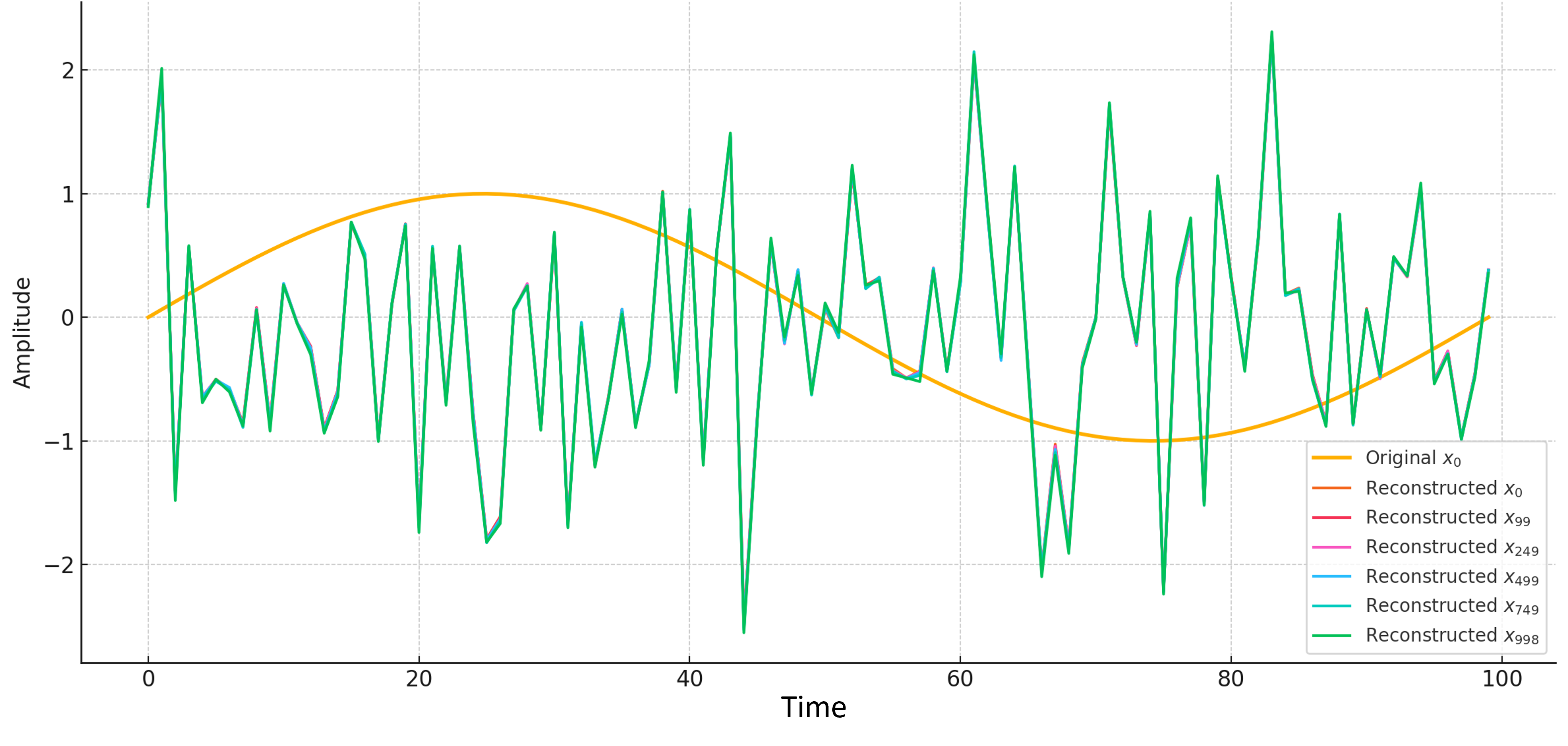}
    \caption{Reverse process (Denoising) in a Simulated EEG signal}
    \label{fig:reverse_process}
\end{figure}

The U-Net architecture consists of encoder and decoder blocks using convolutional and transposed convolutional layers, respectively, each followed by batch normalization and ReLU activations. A critical adaptation of our architecture is the inclusion of temporal embeddings at each step, allowing the model to condition its predictions on the specific noise level associated with each timestep.

The model's training phase leverages mean squared error (MSE) loss to optimize noise prediction:
\begin{equation}
\mathcal{L}(\theta) = \mathbb{E}_{t, x_0, \epsilon} \left[ \left\| \epsilon - \epsilon_\theta \left( \sqrt{\bar{\alpha}_t} x_0 + \sqrt{1 - \bar{\alpha}_t} \epsilon, t \right) \right\|^2 \right]
\end{equation}

We employed the Adam optimizer with an initial learning rate of $1 \times 10^{-4}$ and implemented a learning rate scheduler that reduces the learning rate by half every 20 epochs. Each training session lasted 200 epochs, conducted with a batch size of 32 samples, processed through a dedicated GPU device.

To validate our method, synthetic EEG channels were generated by conditioning on adjacent EEG channels as input, following clearly defined spatial adjacency scenarios. The synthetic signals produced were evaluated using two main metrics: Mean Squared Error (MSE) and Pearson correlation coefficient against original EEG signals. This evaluation ensures that the synthetic data maintain both fidelity and statistical resemblance to genuine EEG recordings.

Our proposed methodology, therefore, offers a robust framework capable of generating realistic EEG data, providing significant potential for applications where EEG data scarcity presents a substantial challenge.

\subsection{Baseline: cWGAN-GP for Synthetic EEG Generation}

To provide a comparative baseline, we implemented a cWGAN-GP, inspired by the methodology presented by \cite{choo2024}. In their study, synthetic motor imagery EEG signals were generated using a cWGAN-GP configured with three convolutional layers in both generator and discriminator networks. The model was trained with the Adam optimizer, using a learning rate of $1 \times 10^{-4}$ and a batch size of 64, under a 5-fold cross-validation protocol. The authors demonstrated that their GAN-based approach was capable of producing EEG signals with sufficient fidelity to enhance the performance of CNN classifiers in motor imagery tasks.

In alignment with this architecture, our implementation of cWGAN-GP followed similar settings, with minor adjustments to input dimensionality and conditioning based on the target channels. This model was used as a benchmark for evaluating the effectiveness of the proposed diffusion-based generation approach. The inclusion of cWGAN-GP as a baseline enables a direct comparison between state-of-the-art adversarial and diffusion-based generative models within the same EEG reconstruction scenario.

\subsection{Classifiers}

In this study, we adopted classifiers that primarily rely on data-driven learning mechanisms, aligning with the methodological guidance provided by our advisor. The goal was to assess models capable of learning patterns directly from the data, without relying on domain-specific heuristics or strong prior assumptions. This ensures a fair comparison between real and synthetic EEG signals, particularly in scenarios where signal characteristics are subtle and data complexity demands flexible yet robust learning strategies.

We selected a diverse set of classifiers to balance simplicity, interpretability, and representational power. KNN and LR serve as baseline methods that evaluate geometric and linear separability, respectively. CNNs were chosen for their ability to extract spatial and temporal features from raw EEG signals. Finally, U-Net was adapted to explore whether its encoder-decoder structure with skip connections could enhance classification by leveraging hierarchical representations. All models were trained and evaluated under consistent conditions to enable direct comparison across real and synthetic datasets.

\subsubsection{K-Nearest Neighbors (KNN)}

The K-Nearest Neighbors classifier was configured with $k=5$ and applied after Z-score standardization of the data. Its objective was to identify the five nearest neighbors in the feature space to perform the classification. Performance was evaluated on both real and synthetic data using accuracy, precision, recall, and F1-Score metrics.

\subsubsection{Logistic Regression (LR)}

LR was used as a baseline linear classifier for comparison, with L2 regularization and optimization via Stochastic Gradient Descent. The model was trained with a limit of 1000 iterations, using normalized data and labels encoded through \textit{LabelEncoder}. The evaluation included accuracy, precision, recall, and F1-Score.

\subsubsection{Convolutional Neural Network (CNN)}

The CNN was employed to capture local patterns in EEG signals through 1D convolutional layers, interleaved with pooling layers and followed by dense layers. ReLU activation functions were used, with a \textit{softmax} activation at the output layer, and the Adam optimizer with a learning rate of $1 \times 10^{-3}$. The architecture was trained for 10 epochs with a batch size of 32.

\subsubsection{U-Net}

The U-Net, traditionally used for biomedical image segmentation tasks, was adapted for EEG classification. The architecture consists of encoding and decoding blocks connected via skip connections, allowing the preservation of spatial information. The model was trained with the same settings as the CNN and evaluated using the same metrics.

\subsection{Evaluation Protocol}

The evaluation of the generated EEG data was conducted using both signal-level and task-level metrics. At the signal level, the Mean Squared Error (MSE) and the Pearson Correlation Coefficient (PCC) between synthetic and real channels were calculated.

For the task-level evaluation, classifiers such as KNN, CNN, and U-Net were trained to discriminate between motor imagery classes. The models were evaluated based on classification accuracy across three data scenarios: only real data, only synthetic data, and mixed datasets. Cross-validation was performed to assess generalization capability.

\section{Experimental Setup}

The experiments were conducted using the BCI Competition III - Dataset V, containing EEG signals recorded from healthy subjects performing motor imagery tasks without feedback. The dataset includes four classes corresponding to imagined movements of the left hand, right hand, feet, and tongue. EEG signals were acquired using 32 channels positioned according to the international 10-20 system, originally recorded at a sampling rate of 512 Hz. No downsampling or resampling procedures were applied; thus, all analyses were performed at the original sampling rate.

The data were segmented into epochs of 1-second duration (512 samples per epoch). For evaluation purposes, the dataset was split into training and testing subsets in a 70\%/30\% ratio, maintaining class balance.

In each experiment, the classification models were trained under two different conditions: (i) using only real EEG data, and (ii) using a hybrid dataset combining real and synthetic EEG channels. The synthetic EEG signals were generated using the previously described DDPM model, reconstructing selected channels based on adjacent electrodes according to spatial proximity criteria established in the literature \cite{michel_2004}.

We evaluated the performance using four different classification models: KNN, LR, CNN, and U-Net. Each model was trained independently with identical hyperparameters and epochs. A 5-fold cross-validation strategy was used, and the average accuracy across folds was computed.

All experiments were implemented in Python 3.11 using PyTorch and MNE-Python libraries. Computations were executed on an NVIDIA A10 GPU with 24GB of memory. The metrics employed for evaluation included Mean Squared Error (MSE), Pearson Correlation Coefficient (PCC), and classification accuracy. Statistical comparisons between real and hybrid datasets were performed using a paired t-test at a significance level of $\alpha = 0.05$.

\subsection{Code and Reproducibility}

To ensure reproducibility and transparency of our experiments, all the code used for preprocessing, training, and evaluating the DDPM and cWGAN-GP models is publicly available at the following anonymous repository in github: \url{https://github.com/halexandre-eng/synthetic-eeg-diffusion-models}

\section{Results and Discussion}

\begin{figure}[htbp]
    \centering
    \includegraphics[width=1\textwidth]{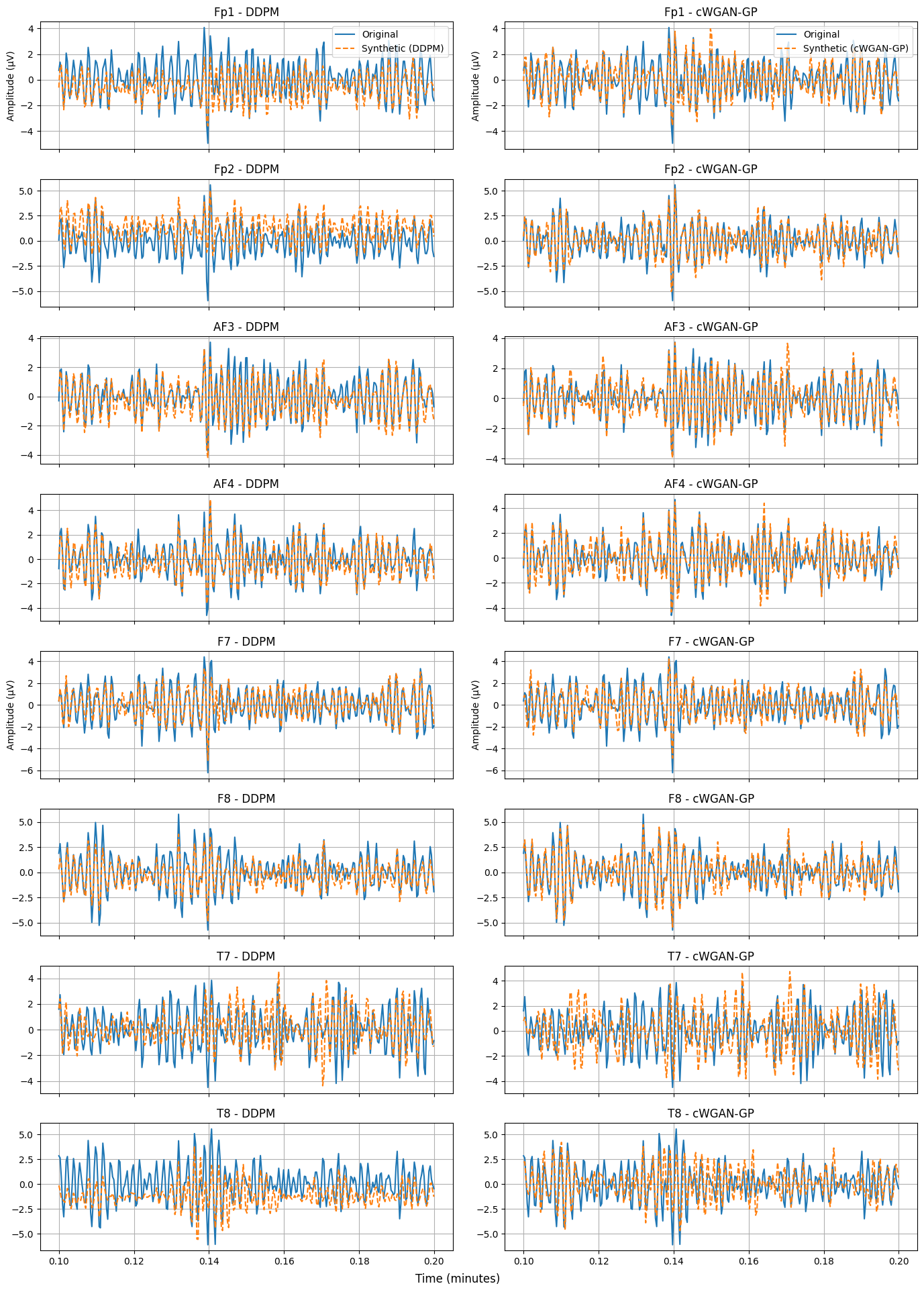}
    \caption{Visual comparison of original and synthetic EEG signals for eight channels. The left column shows signals generated by the DDPM model, while the right column shows results from the cWGAN-GP model. Each plot presents a smoothed 6–12 second interval with real signals in solid lines and synthetic signals in dashed lines.}
    \label{fig:results_ddpm_and_wgan.png}
\end{figure}

The experimental results demonstrate that the synthetic channels generated by both models closely match the real channels, as evidenced by the MSE values and the obtained correlations. Table \ref{tab:avaliacao-canais-difusao} and Figure \ref{fig:results_ddpm_and_wgan.png} present the Mean Squared Error (MSE) and Pearson Correlation metrics for each synthetic channel generated by the diffusion model.

\begin{table}[H]
    \caption{Evaluation of Synthetic Channels by the Diffusion Model}\label{tab:avaliacao-canais-difusao}
    \begin{tabular}{|l|l|l|l|}
    \hline
    \textbf{Target Channel} & \textbf{Input Channels} & \textbf{MSE} & \textbf{Pearson Correlation} \\
    \hline
    AF3 & Fp1, F3 & 5.230499 & 0.791356 \\
    AF4 & Fp2, F4 & 4.655864 & 0.839933 \\
    F7 & FC5, F3 & 4.170372 & 0.854676 \\
    F8 & T8, F4 & 5.457867 & 0.840970 \\
    Fp1 & AF3, F3 & 5.040723 & 0.788673 \\
    Fp2 & AF4, F4 & 11.081029 & 0.683048 \\
    T7 & C3, CP1 & 20.746539 & 0.465758 \\
    T8 & C4, CP2 & 19.659182 & 0.482980 \\
    \hline
    \end{tabular}
\end{table}

Table \ref{tab:avaliacao-canais-wgan_gp} and Figure \ref{fig:results_ddpm_and_wgan.png} present the MSE and PCC metrics for each synthetic channel generated by the cWGAN-GP model.

\begin{table}[H]
    \caption{Evaluation of Synthetic Channels by the cWGAN-GP Model}\label{tab:avaliacao-canais-wgan_gp}
    \begin{tabular}{|l|l|l|l|}
    \hline
    \textbf{Target Channel} & \textbf{Input Channels} & \textbf{MSE} & \textbf{Pearson Correlation} \\
    \hline
    AF3 & Fp1, F3 & 6.206011 & 0.761043 \\
    AF4 & Fp2, F4 & 5.752882 & 0.804417 \\
    F7 & FC5, F3 & 4.954965 & 0.829362 \\
    F8 & T8, F4 & 5.696615 & 0.832832 \\
    Fp1 & AF3, F3 & 5.622387 & 0.760708 \\
    Fp2 & AF4, F4 & 12.132621 & 0.635402 \\
    T7 & C3, CP1 & 29.093567 & 0.369612 \\
    T8 & C4, CP2 & 21.276294 & 0.481955 \\
    \hline
    \end{tabular}
\end{table}

These results show that both models were capable of generating synthetic signals that maintain good correspondence with the real signals, with correlations above 0.63 for most channels and MSE values ranging between 4.17 and 29.09. However, the DDPM consistently achieved lower MSE and higher correlation values across most channels when compared to the cWGAN-GP model. This suggests that the DDPM approach offers better fidelity in reconstructing the temporal structure and statistical properties of the original EEG signals.

\subsection{Evaluation of Generation and Classifier Comparison}

The quality of the generated synthetic signals was evaluated through classification methods to verify whether these data provide sufficient information for motor imagery classification. The implemented classifiers were LR, KNN, CNN, and U-Net. Classification was performed using both the original data and the synthetic data generated by the two models: DDPM and cWGAN-GP.

Table \ref{tab:resultados_classificadores_comparativo} presents the results obtained for the DDPM and cWGAN-GP models.

\begin{table}[H]
\caption{Classifier Results Using Original and Synthetic Data from DDPM and cWGAN-GP}\label{tab:resultados_classificadores_comparativo}
\begin{tabular}{|l|l|l|l|l|l|l|}
\hline
\textbf{Model} & \textbf{Classifier} & \textbf{Dataset} & \textbf{Accuracy} & \textbf{Precision} & \textbf{Recall} & \textbf{F1-Score} \\
\hline
\textbf{Original} & LR & Original (i) & 37.14\% & 18.45\% & 37.14\% & 22.95\% \\
\textbf{Original} & KNN & Original (i) & 93.07\% & 93.20\% & 93.07\% & 93.09\% \\
\textbf{Original} & CNN & Original (i) & 98.89\% & 98.90\% & 98.89\% & 98.89\% \\
\textbf{Original} & U-Net & Original (i) & 98.79\% & 98.79\% & 98.79\% & 98.79\% \\
\hline
\textbf{DDPM} & LR & Synthetic (ii) & 39.24\% & 32.87\% & 39.24\% & 28.88\% \\
\textbf{DDPM} & KNN & Synthetic (ii) & 40.58\% & 42.08\% & 40.58\% & 40.88\% \\
\textbf{DDPM} & CNN & Synthetic (ii) & 95.72\% & 95.75\% & 95.72\% & 95.71\% \\
\textbf{DDPM} & U-Net & Synthetic (ii) & 96.31\% & 96.33\% & 96.31\% & 96.31\% \\
\hline
\textbf{cWGAN-GP} & LR & Synthetic (ii) & 39.82\% & 29.47\% & 39.82\% & 27.82\% \\
\textbf{cWGAN-GP} & KNN & Synthetic (ii) & 41.48\% & 43.02\% & 41.48\% & 41.75\% \\
\textbf{cWGAN-GP} & CNN & Synthetic (ii) & 95.27\% & 95.34\% & 95.27\% & 95.27\% \\
\textbf{cWGAN-GP} & U-Net & Synthetic (ii) & 95.14\% & 95.21\% & 95.14\% & 95.14\% \\
\hline
\end{tabular}
\end{table}

\subsection{Comparison between the Diffusion and cWGAN-GP Models}

Table \ref{tab:comparativo_classificadores} summarizes the comparison of classifier accuracy on the two datasets (Original and Synthetic) for both generative models: Diffusion and cWGAN-GP.

\begin{table}[H]
\caption{Classifier Accuracy Comparison between Diffusion and cWGAN-GP Models}\label{tab:comparativo_classificadores}
\begin{tabular}{|l|l|l|l|}
\hline
\textbf{Classifier} & \textbf{Original (i)} & \textbf{DDPM (ii)} & \textbf{cWGAN-GP (ii)} \\
\hline
LR & 37.14\% & 39.24\% & 39.82\% \\
KNN & 93.07\% & 40.58\% & 41.48\% \\
CNN & 98.89\% & 95.72\% & 95.27\% \\
U-Net & 98.79\% & 96.31\% & 95.14\% \\
\hline
\end{tabular}
\end{table}

It can be observed that both synthetic data generation models, DDPM and cWGAN-GP, show variations in classifier performance compared to the original data. LR exhibited unsatisfactory performance in all cases, indicating that linear methods do not adequately capture the necessary features for motor imagery classification.

The KNN classifier achieved high accuracy on the original data (93.07\%) but showed a significant drop on the synthetic data generated by both models (40.58\% for DDPM and 41.48\% cWGAN-GP), suggesting that KNN is sensitive to differences between real and synthetic data.

Deep learning-based architectures, CNN and U-Net, maintained high performance on the synthetic data. CNN achieved 95.72\% accuracy with the DDPM and 95.27\% with the cWGAN-GP model. U-Net achieved 96.31\% accuracy with the DDPM and 95.14\% with cWGAN-GP. This indicates that the synthetic signals generated provide relevant information for motor imagery classification tasks, with DDPM showing slightly superior performance to the cWGAN-GP in the case of the U-Net.

\subsection{Discussion of Results}

To evaluate the efficacy of the proposed EEG signal generation approach, we conducted comprehensive experiments comparing synthetic data produced by the DDPM and cWGAN-GP models.

The results showed that both models were capable of generating synthetic EEG signals that resembled the original data, as demonstrated by signal-level metrics such as MSE and PCC, as well as task-level classification performance. However, the DDPM consistently achieved lower MSE and higher PCC values across most channels, indicating that it generated signals with greater fidelity and statistical resemblance to the real EEG data.

Regarding classification performance, simpler models such as LR and KNN struggled to maintain accuracy when trained on synthetic data. In particular, KNN — which relies on distance metrics and local neighborhood structures — suffered a significant drop in performance. Even though Z-score normalization was applied uniformly, subtle differences in signal distribution introduced by the generative process, such as variance shifts or minor artifacts, may have distorted the local structures required by KNN, which lacks mechanisms for abstraction or transformation.

In contrast, deep learning-based classifiers such as CNN and U-Net performed well across both real and synthetic datasets. These models benefit from hierarchical feature extraction and demonstrate generalization capabilities, even when trained on partially synthetic data. This reinforces the relevance of expressive models in pipelines that integrate generative synthetic EEG.

While both DDPM and cWGAN-GP produced viable synthetic data, the diffusion-based model outperformed cWGAN-GP in signal-level evaluations, offering more reliable reconstructions. Although classification metrics were similar between models—particularly in CNN and U-Net—the consistently higher signal quality of DDPM suggests that it is a more suitable choice for EEG reconstruction tasks.

These findings support the use of generative models, especially diffusion-based ones, to enhance EEG data availability and improve the performance of machine learning systems in brain-computer interface applications.

\section{Conclusion}

In this work, we presented an approach for the generation of synthetic EEG signals aimed at motor imagery tasks using a DDPM. The proposed method demonstrated performance in terms of signal fidelity and classification accuracy when compared to a traditional generative model, the cWGAN-GP. 

Although both models produced synthetic EEG signals suitable for downstream classification, the DDPM approach achieved lower MSE and higher PCC across most channels, indicating a closer resemblance to real EEG signals. While classification metrics were comparable for both models—particularly in deep learning-based classifiers—DDPM demonstrated greater consistency in signal-level quality, making it a more reliable method for EEG reconstruction.

Our results confirm that diffusion models are capable of effectively reconstructing missing EEG channels and providing realistic data for augmenting training datasets. The combination of real and synthetic data led to improved model generalization, highlighting the role of synthetic EEG in overcoming data scarcity challenges in BCI applications.

This study contributes to the literature by demonstrating the viability and robustness of DDPM-based synthetic EEG generation for channel reconstruction, validated through multi-classifier evaluations.

Despite these promising outcomes, we acknowledge significant limitations such as the relatively small dataset size and the lack of EEG-specific evaluation metrics (e.g., spectral power and phase synchronization). Therefore, future analyses incorporating larger and more diverse datasets and EEG-specific metrics will be essential to further validate and enhance the robustness and generalizability of our findings.

As future work, we propose extending this methodology to datasets involving multiple subjects, incorporating temporal context through transformer-based architectures, and evaluating the effectiveness of synthetic EEG data in real-time BCI scenarios. Additionally, it would be valuable to explore the generation of a greater number of synthetic channels—gradually increasing the number of replaced channels—to determine the maximum feasible number that can be synthetically reconstructed without compromising the performance of downstream classification models.

%
%
\bibliographystyle{splncs04}
\bibliography{samplepaper}

\begin{thebibliography}{10}
\providecommand{\url}[1]{\texttt{#1}}
\providecommand{\urlprefix}{URL }
\providecommand{\doi}[1]{https://doi.org/#1}

\bibitem{alotaiby2015review}
Alotaiby, T.N., Alshebeili, S.A., Alshawi, T., Ahmad, I., Abd El-Samie, F.E.: A review of channel selection algorithms for eeg signal processing. EURASIP Journal on Advances in Signal Processing  \textbf{2015},  1--21 (2015)

\bibitem{cai2025}
Cai, Y., Meng, Z., Huang, D.: Dhct-gan: Improving eeg signal quality with a dual-branch hybrid cnn–transformer network. Sensors  \textbf{25}(1) (2025). \doi{10.3390/s25010231}, \url{https://www.scopus.com/inward/record.uri?eid=2-s2.0-85214510779&doi=10.3390%2fs25010231&partnerID=40&md5=a9777d1d21a45fae5f0a10bad8e365c0}, cited by: 2; All Open Access, Gold Open Access

\bibitem{casson_2010}
Casson, A.J., Yates, D.C., Smith, S.J., Duncan, J.S., Rodriguez-Villegas, E.: {Wearable Electroencephalography}. {IEEE Engineering in Medicine and Biology Magazine}  \textbf{29}(3),  44--56 (2010)

\bibitem{choo2024}
Choo, S., Park, H., Jung, J.Y., Flores, K., Nam, C.S.: Improving classification performance of motor imagery bci through eeg data augmentation with conditional generative adversarial networks. Neural Networks  \textbf{180} (2024). \doi{10.1016/j.neunet.2024.106665}, \url{https://www.scopus.com/inward/record.uri?eid=2-s2.0-85203163358&doi=10.1016%2fj.neunet.2024.106665&partnerID=40&md5=2ae778b809f1c293a13260d7e5232454}, cited by: 1

\bibitem{dhariwal_2021}
Dhariwal, P., Nichol, A.: {Diffusion Models Beat GANs on Image Synthesis}. In: {Advances in Neural Information Processing Systems}. vol.~34, pp. 8780--8794 (2021)

\bibitem{goodfellow_2014}
Goodfellow, I.J., Pouget-Abadie, J., Mirza, M., Xu, B., Warde-Farley, D., Ozair, S., Courville, A., Bengio, Y.: {Generative Adversarial Nets}. In: {Advances in Neural Information Processing Systems}. vol.~27 (2014)

\bibitem{ho_2020}
Ho, J., Jain, A., Abbeel, P.: {Denoising Diffusion Probabilistic Models}. In: {Advances in Neural Information Processing Systems}. vol.~33, pp. 6840--6851 (2020)

\bibitem{ho2020denoising}
Ho, J., Jain, A., Abbeel, P.: Denoising diffusion probabilistic models. Advances in Neural Information Processing Systems (NeurIPS)  \textbf{33},  6840--6851 (2020)

\bibitem{luo2020}
Luo, J., Zhu, F., Li, W.: Eeg signal enhancement using gans for emotion recognition. IEEE Transactions on Affective Computing  \textbf{11}(4),  676--687 (2020)

\bibitem{michel_2004}
Michel, C.M., Murray, M.M., Lantz, G., Gonzalez, S., Spinelli, L., Grave~de Peralta, R.: {EEG Source Imaging}. {Clinical Neurophysiology}  \textbf{115}(10),  2195--2222 (2004)

\bibitem{penava_2023}
Penava, S., Radivojević, A., Šćepanović, D., Milić, D., Milovanović, B.: {A Novel Approach for EEG Data Augmentation Using Generative Adversarial Networks}. {Neural Networks}  \textbf{157},  263--274 (2023)

\bibitem{qian2024}
Qian, D., Zeng, H., Cheng, W., Liu, Y., Bikki, T., Pan, J.: Neurodm: Decoding and visualizing human brain activity with eeg-guided diffusion model. Computer Methods and Programs in Biomedicine  \textbf{251} (2024). \doi{10.1016/j.cmpb.2024.108213}, \url{https://www.scopus.com/inward/record.uri?eid=2-s2.0-85192974953&doi=10.1016%2fj.cmpb.2024.108213&partnerID=40&md5=4640d777b9cf8f8daedd651237952646}, cited by: 2

\bibitem{raoof2023}
Raoof, A., Gupta, R.: Conditional gan for spatio-temporal eeg signal generation in motor imagery. In: Proceedings of the 45th Annual International Conference of the IEEE Engineering in Medicine and Biology Society (EMBC). pp. 1234--1238. IEEE (2023)

\bibitem{sohl_2015}
Sohl-Dickstein, J., Weiss, E., Maheswaranathan, N., Ganguli, S.: {Deep Unsupervised Learning Using Nonequilibrium Thermodynamics}. In: {Proceedings of the 32nd International Conference on Machine Learning}. pp. 2256--2265 (2015)

\bibitem{song_2020}
Song, Y., Ermon, S.: {Score-Based Generative Modeling Through Stochastic Differential Equations}. In: {International Conference on Learning Representations} (2020)

\bibitem{tang2022channel}
Tang, J., Chen, L., Xu, M., Shu, X., Hu, X., Xu, J.: Channel selection for motor imagery eeg based on sequential backward floating search. Frontiers in Neuroscience  \textbf{16},  1045851 (2022)

\bibitem{torma2023}
Torma, L., Varga, T., Kovács, K.: Eeg signal generation using diffusion probabilistic models. In: Proceedings of the 31st European Signal Processing Conference (EUSIPCO). pp. 987--991. IEEE (2023)

\bibitem{yu2020cross}
Yu, J., Yu, Z.: Cross-correlation based discriminant criterion for channel selection in motor imagery bci systems. arXiv preprint arXiv:2012.01749  (2020)

\bibitem{zhao2023eeg}
Zhao, M., Zhang, S., Mao, X., Sun, L.: Eeg topography amplification using fastgan-asp method. Electronics  \textbf{12}(24), ~4944 (2023)

\end{thebibliography}

\end{document}